\documentclass[preprint,aps,showkeys,12pt]{revtex4}
 \usepackage{graphicx}
\usepackage[pdftex]{color}
\usepackage{amsmath}
\usepackage{amssymb}
\usepackage{wasysym}

\newtheorem{theorem}{Theorem}[section]

\newtheorem{definition}[theorem]{Definition}

\def \beq{\begin{equation}}
\def \eeq{\end{equation}}
\def \beqa{\begin{eqnarray}}
\def \eeqa{\end{eqnarray}}
\def \beqan{\begin{eqnarray*}}
\def \eeqan{\end{eqnarray*}}
\def \bea{\begin{eqnarray}}
\def \eea{\end{eqnarray}}
\def\con{{\rm con}\ }
\def\aff{{\rm aff}\ }

\def\ext{{\rm ext}\ }

\def \proof {\noindent {\em Proof. }}

\newcommand{\R}{\mathbb{R}}

\newcommand{\Prob}{\mathbb{P}}

\newcommand{\E}{\mathbb{E}}

\newcommand{\qed}{\hfill $\Box$ \vskip 2ex}

\def \proof {\noindent {\em Proof. }}
\def \qed{\hfill $\Box$ \vskip 2ex}

\begin{document}

\title{A note on the geometric interpretation of  Bell's inequalities}
\author{Paolo Dai Pra, Michele Pavon, Neeraja Sahasrabudhe}\thanks{
The work of M. Pavon was partially supported by  the QuantumFuture research grant of the University of Padova and by an Alexander von Humboldt Foundation fellowship at the  Institut f\"{u}r Angewandte Mathematik, Universit\"{a}t Heidelberg, Germany.}
\affiliation{Dipartimento di Matematica, Universit\`a di Padova, via
Trieste 63, 35121 Padova, Italy}

%
%
\date{\today}

\begin{abstract}
Using results of Pitowsky and Gupta, we show in a direct, elementary fashion that, in the case of three spins, Bell's inequalities indeed provide a representation of the tetrahedron of all spin correlation matrices as intersection of half-spaces.\\\\Mathematics Subject Classification[2010]:52B12,62H20,81P40
\end{abstract}


\keywords{Bell's inequalities, correlation matrix}

\maketitle


\section{Introduction}
The non-local character of quantum correlations is manifested by the violation of Bell's
inequality \cite{Bell}, and, more generally, of the so-called CHSH inequality \cite{CHSH,Cir}. The experimental violation  of these inequalities \cite{ASPECT_GRANGIER_ROGER} 
constitutes by now an important part of modern scientific culture. This property and its relation to entanglement has in fact become one of the cornerstones of quantum information theory, see e.g. \cite{WW}. The relation between Bell's inequalities and convex geometry is also well-known: It has been studied somewhat independently in the mathematical physics literature \cite{Pitowsky1991}, \cite[Appendix A]{WW}, \cite{Pitowsky2001}, \cite{Pitowsky2008} and in the statistics literature  \cite{Balasubramanian_Gupta_Parthasarathy,JCG}. In particular, Bell's inequalities are known to be related to the representation of a polytope as intersection of  halfspaces. Nevertheless, the problem of finding a minimal set of Bell-like inequalities is in general still open as it relates to the hard problem of finding all the facets of a polytope \cite{Pitowsky1991,Pitowsky2001,WW}. 

\begin{figure}[h!]
\begin{center}
\includegraphics[width=4cm]{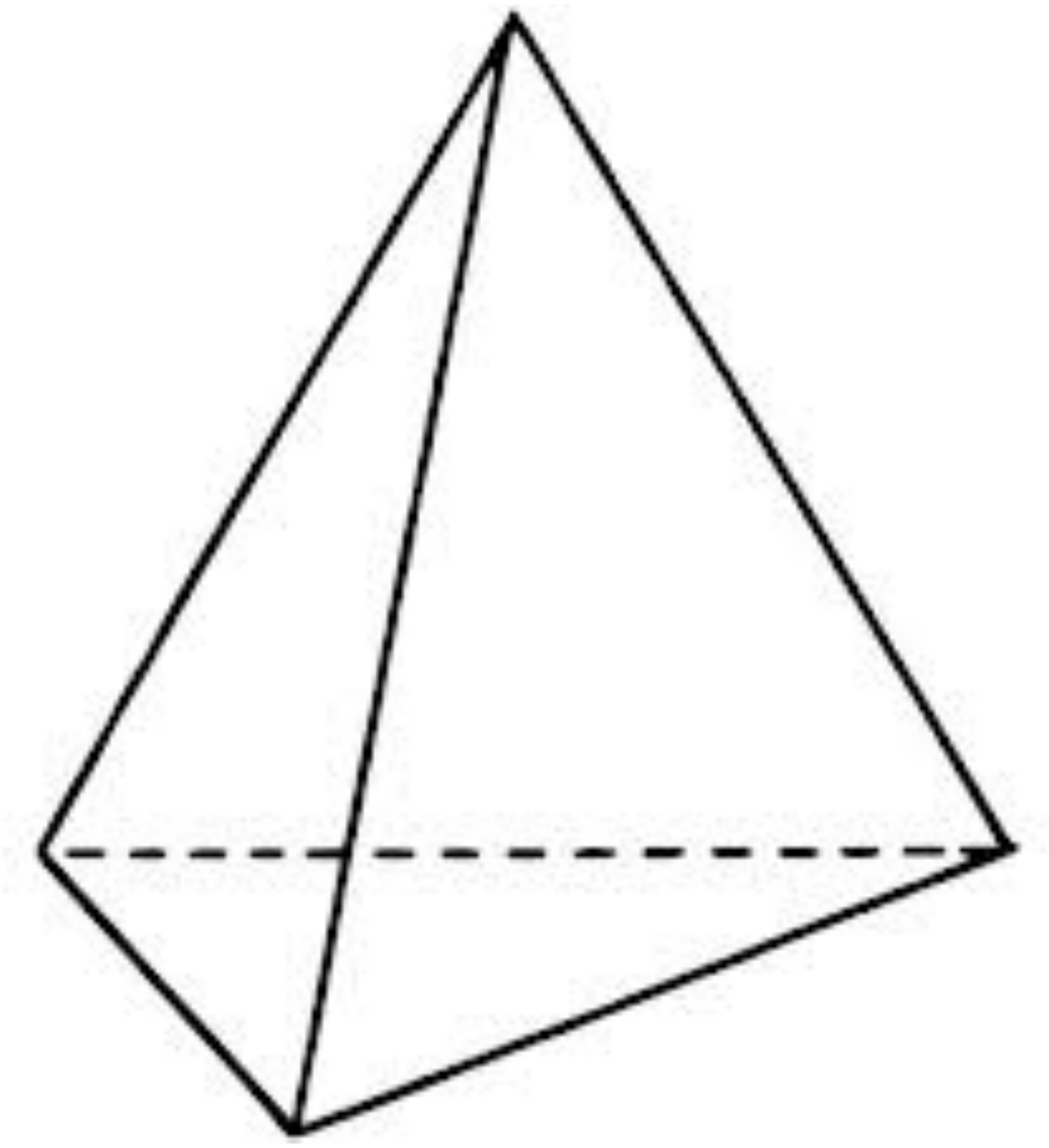}
\end{center}
\caption{Tetrahedron} 
\label{TETRAHEDRON}
\end{figure}

The purpose of this note is to show that, in the simple but important original case of $n=3$ spin random variables, the fact that Bell's inequalities provide a facets representation of the polytope of all spin correlation matrices can be proven in a direct, straightforward way using the available description of the extreme points \cite{Pitowsky1991,JCG}. This elementary derivation, which is apparently missing in the large body of literature on this topic, only works in the $n=3$ case. In such a case, and for no other $n\ge 3$, the polytope is actually a simplex, so that it constitutes a {\em tetrahedron}, see Figure \ref{TETRAHEDRON}. 

\section{Background}
Let $E$ be a $p$-dimensional real vector space. We denote by $\aff A$ the {\em affine hull} of $A\subseteq E$, namely the intersection of all linear manifolds in $E$ containing $A$. It can be shown that
$$\aff A=\{\sum_{i=1}^m\lambda_ix^i | x^i\in A, \sum_{i=1}^m\lambda_i=1, m \;{\rm finite}\}.
$$
We denote by $\con A$ the {\em convex hull} of $A$, namely the intersection of all convex sets containing $A$. It is easy to see that
$$\con A=\{\sum_{i=1}^m\lambda_ix^i | x^i\in A, \sum_{i=1}^m\lambda_i=1, \lambda_i\ge 0\}.
$$
\begin{definition} A convex set $D\subseteq E$ is a $d$-{\em simplex} if $D=\con(x_0,x_1,\ldots,x_d)$ and the points $x_0,x_1,\ldots,x_d$ are {\em affinely independent}, i.e. $x_i\not\in\aff(x_0,x_1,\ldots,x_{i-1},x_{i+1},\ldots,x_d), \forall i$. Necessarily, $d\le p$.
\end{definition}
If $D$ is a $d$-{\em simplex}, every point $x\in D$ admits a {\em unique} representation as convex combination of the points $x_0,x_1,\ldots,x_d$. Moreover, $\ext D$, the set of {\em extreme points} of $D$, is indeed $\ext D=\{x_0,x_1,\ldots,x_d\}$.
\begin{definition} A set $P\subseteq E$ which is the convex hull of a finite number of points is called a {\em (convex) polytope}. 
\end{definition}

Consider a family of spin random variables $\{\xi_i,1\le i\le n\}$ with  $\Prob(\xi_i=-1)=\Prob(\xi_i=1)=1/2$ and form the corresponding column vector $\xi$, where $\xi^T=(\xi_1,\ldots,\xi_n)$, $T$ denoting transposition.  The random vector $\xi$ has zero expectation and  takes values in $\Omega_n=\{-1,1\}^n$. Let 
$$\Sigma=\E\{\xi\xi^T\}=\left(\sigma_{ij}\right)_{i,j=1}^n$$ 
denote the corresponding covariance/correlation matrix. Given the $\sigma_{ij}$, the question whether the corresponding matrix can be realized as the correlation matrix of spin random variables is a nontrivial one. Indeed, while any symmetric, positive semidefinite matrix $\Sigma$ can be realized as the covariance of a Gaussian random vector, the conditions on $\Sigma$ become much more stringent if we require realization through spin  random variables. Since \cite{Bell}, it is known that for three spins variables, the $\sigma_{ij}$ need to satisfy Bell's inequalities to be realizable. Actually, it was shown in \cite{Balasubramanian_Gupta_Parthasarathy} that, in the case $n=3,4$, Bell's inequalities 
\begin{equation}\label{Bellsinequalities}
1+\epsilon_i\epsilon_j \sigma_{ij}+ \epsilon_i\epsilon_k \sigma_{ik} + \epsilon_j\epsilon_k \sigma_{jk}\geq 0, \quad\forall  \quad1\le i<j<k\le n,\quad
\epsilon_i\in\{-1,\,+1\},
\end{equation}
are necessary and sufficient for the matrix $\Sigma=\E\{\xi\xi^T\}$ to be the correlation matrix of $n$ spin random variables.  Notice that a correlation matrix is symmetric and has ones on the main diagonal. It is therefore determined by the elements above the main diagonal. Hence, these matrices are in one to one correspondence with a $n(n-1)/2$-dimensional Euclidean space. For  $n\ge 2$,  let ${\mathcal C}_n$ denote the set of correlation matrices  that can be realized through spin random variables.
Pitowsky \cite{Pitowsky1991} and Gupta \cite[Theorem3.1]{JCG} showed that ${\mathcal C}_n$ is a {\em polytope}. It consists namely of all convex combinations of a finite number of correlation matrices. They also gave an explicit description of the extreme points (matrices) of this set. These are rank-$1$ matrices of the form $\omega^{(n)}(\omega^{(n)})^T$, where the $n$-dimensional  vector $\omega^{(n)}$ has entries equal to $1$ or $-1$. Notice that $\omega^{(n)}$ and $-\omega^{(n)}$ generate the same correlation matrix so that there are $2^{n-1}$ different such matrices. Since a polytope simply consists  of all convex combinations of its extreme points, this completely characterizes ${\mathcal C}_n$.

 \section{Geometric meaning of Bell's inequalities}
 
By the Minkowski-Weyl theorem \cite{Z},  \cite[Appendix A]{WW}, a polytope, besides being the convex hull of its extreme points (V-representation), can also be dually described as the intersection of  half-spaces through a system of linear inequalities (H-representation). 
For instance, a triangle in the plane is the intersection of three half-planes determined by the straight lines containing its sides. It is known \cite{Pitowsky2001,WW} that the latter description for ${\mathcal C}_n$ is related to the Bell's inequalities (\ref{Bellsinequalities}).  Consider now the $n=3$ case, where (\ref{Bellsinequalities}) take the form
\begin{eqnarray}\label {Bell1}1+ \sigma_{12} +\sigma_{13}+\sigma_{23}\ge 0,\\ 1- \sigma_{12} +\sigma_{13}-\sigma_{23}\ge 0,\\ 1+ \sigma_{12} -\sigma_{13}-\sigma_{23}\ge 0,\\ 1- \sigma_{12} -\sigma_{13}+\sigma_{23}\ge 0.\label{Bell4}
\end{eqnarray}
\begin{theorem} The set ${\mathcal C}_3$ is a $3$-simplex, namely a tetrahedron. Bell's inequalities (\ref{Bell1})-(\ref{Bell4}) provide the H-representation of the tetrahedron ${\mathcal C}_3$. Thus, any element in ${\mathcal C}_3$ satisfies Bell's inequalities (\ref{Bell1})-(\ref{Bell4}). Conversely, any correlation matrix  satisfying Bell's inequalities belongs to ${\mathcal C}_3$. 
\end{theorem}
\proof
Consider the four vectors
$$v_1=\left(\begin{matrix}+1\\+1\\+1\end{matrix}\right),\quad v_2=\left(\begin{matrix}+1\\-1\\+1\end{matrix}\right),\quad v_3=\left(\begin{matrix}+1\\+1\\-1\end{matrix}\right),\quad  v_4=\left(\begin{matrix}-1\\+1\\+1\end{matrix}\right).
$$
By \cite{Pitowsky1991,JCG}, the extreme points of  ${\mathcal C}_3$  are given by the four matrices 
\begin{eqnarray*}
&&\Sigma_1=v_1v_1^T=\left[\begin{matrix}+1& +1& +1\\ +1& +1 & +1\\ +1 & +1 & +1\end{matrix}\right],\quad \Sigma_2=v_2v_2^T=\left[\begin{matrix}+1& -1& +1\\ -1& +1 & -1\\ +1 & -1 & +1\end{matrix}\right],\\
&&\Sigma_3=v_3v_3^T=\left[\begin{matrix}+1& +1& -1\\ +1& +1 & -1\\ -1 & -1 & +1\end{matrix}\right],\quad\Sigma_4=v_4v_4^T=
\left[\begin{matrix}+1& -1& -1\\ -1& +1 & +1\\ -1 & +1 & +1\end{matrix}\right].
\end{eqnarray*}
Consider now a symmetric matrix $\Sigma$ with ones on the main diagonal. As observed before, they can be bijectively mapped onto a $3$-dimensional space. Let
$$A:=\left[\begin{matrix}+1& +1& +1&+1\\ +1& -1 & +1&-1\\ +1 & +1 & -1&-1\\+1 & -1 & -1&+1\end{matrix}\right],\quad X=\left[\begin{matrix}1\\ \sigma_{12}\\ \sigma_{13}\\ \sigma_{23}\end{matrix}\right].
$$
Knowledge of $X$ is equivalent to knowledge of $\Sigma$. Define
$$Y=AX=\left[\begin{matrix}1+ \sigma_{12} +\sigma_{13}+\sigma_{23}\\ 1- \sigma_{12} +\sigma_{13}-\sigma_{23}\\ 1+ \sigma_{12} -\sigma_{13}-\sigma_{23}\\ 1- \sigma_{12} -\sigma_{13}+\sigma_{23}\end{matrix}\right].
$$
Then Bell's inequalities (\ref{Bell1})-(\ref{Bell4}) simply state that the vector $Y$ belongs to the positive orthant $ Y\in\R^4_+$, it has namely all nonnegative components.
Consider a symmetric, $3\times 3$  matrix $\Sigma$ with ones on the main diagonal expressed as {\em linear} combination of the $\Sigma_i$:
\begin{equation}\label{form}\Sigma=\lambda_1 \Sigma_1 + \lambda_2 \Sigma_2 + \lambda_3\Sigma_3 + \lambda_4\Sigma_4,\quad  \lambda_i\in\R.
\end{equation}
Since $\sigma_{ii}=1$ and the $\Sigma_i, i=1,2,3,4$ have the same property, it follows that necessarily
\begin{equation}\label{sum}
\sum_{i=1}^4\lambda_i=1,
\end{equation}
namely $\Sigma$ is in $\aff\{ \Sigma_1, \Sigma_2,\Sigma_3,\Sigma_4\}$. From (\ref{form}), a simple calculation yields
$$X=\left[\begin{matrix}1\\ \sigma_{12}\\ \sigma_{13}\\ \sigma_{23}\end{matrix}\right]=A\left[\begin{matrix}\lambda_1\\ \lambda_2\\ \lambda_3\\ \lambda_4\end{matrix}\right].
$$
Observe that $\frac{1}{2}A$ is involutory, namely $A^2=4 I_4$. Since $A$ is invertible, it follows that any symmetric matrix with ones on the main diagonal may be expressed as in (\ref{form}). In particular, every element in the convex hull of $\{\Sigma_1,\Sigma_2,\Sigma_3,\Sigma_4\}$ admits a unique representation. Hence, the $\Sigma_i, i=1,2,3,4$ are affinely independent and generate a simplex, namely a tetrahedron. Moreover, we have
$$Y=AX=A^2\left[\begin{matrix}\lambda_1\\ \lambda_2\\ \lambda_3\\ \lambda_4\end{matrix}\right]=4\left[\begin{matrix}\lambda_1\\ \lambda_2\\ \lambda_3\\ \lambda_4\end{matrix}\right].
$$
Thus Bell's inequalities are {\em equivalent} to having $\lambda_i\ge 0, i=1,2,3,4$. Taking (\ref{form}) and (\ref{sum}) into consideration, we conclude that a convex combination of the $\Sigma_i$ is a correlation matrix satisfying Bell's inequalities. Namely, any element in ${\mathcal C}_3$ satisfies Bell's inequalities (\ref{Bell1})-(\ref{Bell4}). Conversely, any correlation matrix  satisfying Bell's inequalities belongs to the convex hull of the $\Sigma_i$, i.e. it belongs to ${\mathcal C}_3$. 
\qed

\section{Closing comments} 
The above elementary argument does not extend to higher dimensions. In particular, ${\mathcal C}_n$ is not a simplex for $n>3$. For instance, we know from \cite{Balasubramanian_Gupta_Parthasarathy} that also in the case $n=4$ Bell's inequalities are necessary and sufficient for $\Sigma$ to be realizable as correlation matrixof spin random variables. There are $16$ such  inequalities. The set ${\mathcal C}_4$ has eight extreme points whereas $\Sigma$ is determined by six elements (those above the main diagonal). But, in a $6$-dimensional space, a simplex can have at most $7$ extreme points. Thus the polytope ${\mathcal C}_4$ is not a simplex.  More generally, the equation 
$$2^{n-1}=\frac{n(n-1)}{2}+1$$
relating the number of extreme points of ${\mathcal C}_n$ and the maximum number of affinely independent points in a space of dimension $n(n-1)/2$ has no solution for $n>3$.
Moreover, for $n\ge 5$,  Bell's inequalities (\ref{Bellsinequalities}) are known not to be sufficient for a correlation matrix to be  the correlation matrix of spin random variables \cite{JCG}.


\begin{thebibliography}{1}

\newcommand{\AUTHORS}[1] {{ #1}}
\newcommand{\TITLE}  [1] {{\em #1}}
\newcommand{\VOLUME} [1] {{vol. {\bf #1}}}
\newcommand{\PAGES}  [1] {{pp. #1}}

\newcommand{\BOOK} [5] {\AUTHORS{#1}, \TITLE{#2}, #3, #4, #5}
\newcommand{\PAPER}[6] {\AUTHORS{#1}, \TITLE{#2}, #3 \VOLUME{#4}, \PAGES{#5}, #6}
\newcommand{\CONF}[5] {\AUTHORS{#1}, \TITLE{#2}, #3 \PAGES{#4}, #5}
\newcommand{\ARXIV}[3] {\AUTHORS{#1}, \TITLE{#2}, arXiv e-print {\tt #3}}

	

	
\bibitem{ASPECT_GRANGIER_ROGER}
\PAPER
 {A. Aspect, A. P. Grangier and G. Roger}
 {Experimental tests of realistic local theories via Bell's theorem}
{Physical Review Letters}
{47}
{460-463}
{1981}.
		



	

	
	

\bibitem{Balasubramanian_Gupta_Parthasarathy}
\PAPER
{K. Balasubramanian, J.C. Gupta and K.R. Parthasarathy}
{Remarks on Bell's Inequalities for Spin Correlations}
{Sankhya: The Indian Journal of Statistics}
{60}
{29-35}
{1998}.
	

	
\bibitem{Bell}
\PAPER
{J. S. Bell}
{On the Einstein Podolsky Rosen Paradox}
{Physics}
{1}
{195-200}
{1964}.

\bibitem{Cir}
\PAPER
{B. S. Cirel\'{}son}
{Quantum generalizations  of Bell's inequality}
{Lett. Math. Phys.} 
{4} 
{93-100}
{1980}.

\bibitem{CHSH}
\PAPER
{J. F. Clauser, M. A. Horne, A. Shimony and R. A. Holt}
{Phys. Rev. Lett.} 
{23}
{880-884}
{1969}.	
	
\bibitem{JCG}
\PAPER
{J. C. Gupta}
{Characterisation of Correlation Matrices of Spin Variables}
{Sankhya: The Indian Journal of Statistics}
{61}
{282-285}
{1999}.

\bibitem{Pitowsky1991}
\PAPER
{I. Pitowsky}
{Correlation Polytopes: Their Geometry and Complexity} 
{Mathematical Programming} 
{50} 
{395-414}
{1991}. 


\bibitem{Pitowsky2001}
\PAPER
{I. Pitowsky and K. Svozil}
{Optimal tests of quantum nonlocality} 
{Physical Review A}
{64}
{014102} 
{2001}.

\bibitem{Pitowsky2008}
\PAPER
{I. Pitowsky}
{On the geometry of quantum correlations} 
{Physical Review A} 
{77} 
{062109-1-5 }
{2008}. 


\bibitem{WW}
\PAPER
{R. F. Werner and  M. M. Wolf}
{Bell inequalities and entanglement}
{Quant. Inform. Comput.} 
{1} 
{1-25}
{2001}.
	
\bibitem{Z}
\BOOK
{G. Ziegler}
{Lectures on polytopes}
{Springer-Verlag}
{New York}
 {1997}.


	
	
\end{thebibliography}
\end{document}